# A Search for the Higgs Boson in H → ZZ → 2 leptons 2 jets Mode

Ashish Kumar
*The State University of New York at Buffalo, NY 14260, USA*
(on behalf of the CMS Collaboration)

A search for the Standard Model Higgs boson decaying to two Z bosons with subsequent decay to a final state with two leptons and two quark-jets, H → ZZ → $2\ell$ 2jets, where $\ell$ = e, μ, is discussed. It is based on an integrated luminosity of 1 fb$^{-1}$ of proton-proton collision data at center of mass energy √s = 7 TeV collected by the CMS experiment at the LHC. The data are compared to the expected Standard Model backgrounds. No evidence for the Higgs boson is found and upper limits are placed on its production cross section over the range of masses between 200 GeV and 600 GeV.

## 1. Introduction

The search for the Standard Model (SM) Higgs boson is one of the most important goals of the experiments at the Large Hadron Collider (LHC). Direct searches at the CERN LEP $e^+e^-$ collider have set a lower limit of 114.4 GeV on the Higgs boson mass, $m_H$, at 95% confidence level (CL) [1]. Searches by the CDF and D0 experiments at the Fermilab Tevatron collider have explored the mass range up to 200 GeV and exclude the additional region 156 < $m_H$ < 177 GeV [2]. For $m_H$ greater than twice the Z boson mass, $m_Z$, a significant fraction of Higgs bosons decay to two Z bosons. The H→ZZ → $4\ell$ channel, where $\ell$ represents charged leptons e and μ, is well known to offer excellent separation between backgrounds and a potential Higgs signal, but is limited in statistics due to the small branching fraction for both Z bosons to decay into either electrons or muons. In this contribution, we present results of a search for the SM Higgs boson by the CMS experiment where the Higgs boson decays to two Z bosons with a subsequent decay to two leptons and two quark jets, H→ZZ→$2\ell$ 2jets [3]. The branching fraction of this decay channel is about 20 times higher than that of H→ZZ→$4\ell$. This may lead to a better sensitivity to the SM Higgs boson production at higher masses, where background can be effectively suppressed by kinematic selection. The main sources of background in this channel are Z+jets and a small contribution of top-pair and electroweak diboson production. To suppress these backgrounds, the searches are performed assuming the Higgs boson decays to two on-shell Z bosons. This effectively limits the search sensitivity to Higgs boson masses above twice the mass of the Z boson. The Higgs boson mass range considered in the search presented here is between 200 and 600 GeV, a range extending beyond the sensitivities the Tevatron experiments. The analysis discussed here is based on proton-proton collision data at the center of mass energy √s = 7 TeV, collected by the CMS experiment at the LHC and corresponding to an integrated luminosity of 1 fb$^{-1}$. Similar search in this channel has been performed by the ATLAS experiment utilizing 35 pb$^{-1}$ of data [4].

## 2. Event Selection

The H→ZZ→$2\ell$ 2jets events are characterized by two high transverse momentum ($p_T$) electrons or muons consistent with the decay of a Z boson and a pair of jets. A detailed description of the CMS detector can be found elsewhere [5]. Muons are measured with the all-silicon tracker and the muon system. Electrons are detected in the electromagnetic calorimeter as energy clusters matched with tracks in the tracker. Both, muons and electrons are required to be isolated and to pass quality requirements. The details of electron and muon identification criteria are described elsewhere [6]. The leading lepton must have $p_T$ > 40 GeV and the other one $p_T$ > 20 GeV. They are measured in the pseudorapidity range |η| < 2.4 for muons, and |η| < 2.5 for electrons. Jets are reconstructed with the Particle Flow (PF) algorithm [7], which is a technique with the aim of reconstructing all particles produced in a given collision event through the combination of information from all sub-detectors. The reconstructed particle candidates are clustered to form PF jets with the anti-$k_T$ algorithm with radius R = 0.5 [8]. Jets with $p_T$ > 30 GeV and |η| < 2.5 are considered in the analysis and those which overlap with isolated leptons within ΔR = 0.5 are removed.

In order to further reduce the background, the dijet and dilepton invariant masses are required to be within 75 GeV < $m_{jj}$ < 105 GeV and 70 GeV < $m_{\ell\ell}$ < 110 GeV, respectively. Further kinematic selection exploits five angular observables that fully describe the kinematics of the gg→H→ZZ→$2\ell$ 2jets process. They contain most of the discriminating power between signal and background and an angular likelihood discriminant is constructed based on the probability ratio of the signal and background hypotheses, as described in the Ref. [9]. Figure 1(a) shows the distribution of angular likelihood discriminant. A powerful handle in the signal versus background discrimination is also offered by the parton flavor of the jets. Jets in signal events originate from the hadronization of quarks, while, gluon radiation plays a major role in the dominant background from Z + jets. Therefore, the main features that discriminate signal from background are the relatively large contribution of heavy flavor quarks and the absence of gluons. These features are exploited in the analysis by performing tagging of the b-flavor and introducing a likelihood discriminant which separates gluon and light-quark jets on a statistical basis. To identify jets originating from the hadronization of bottom quarks, Track Counting High Efficiency (TCHE) b-tagging algorithm is used [10]. The data are split into three exclusive b-tag categories containing zero, one and two b-tagged jets (Fig. 1(b)). The composition of the expected signal and background varies significantly among the three categories. The "0 b-tag" category is dominated by the Z+jets background, and from these events, a "gluon-tagged" category is selected, which is then excluded in further analysis. To help suppress the top-pair background in the category with "2-btag", a cut on missing transverse energy (MET) significance is used : the top events are expected to have a large MET due to presence of two neutrinos in the event.

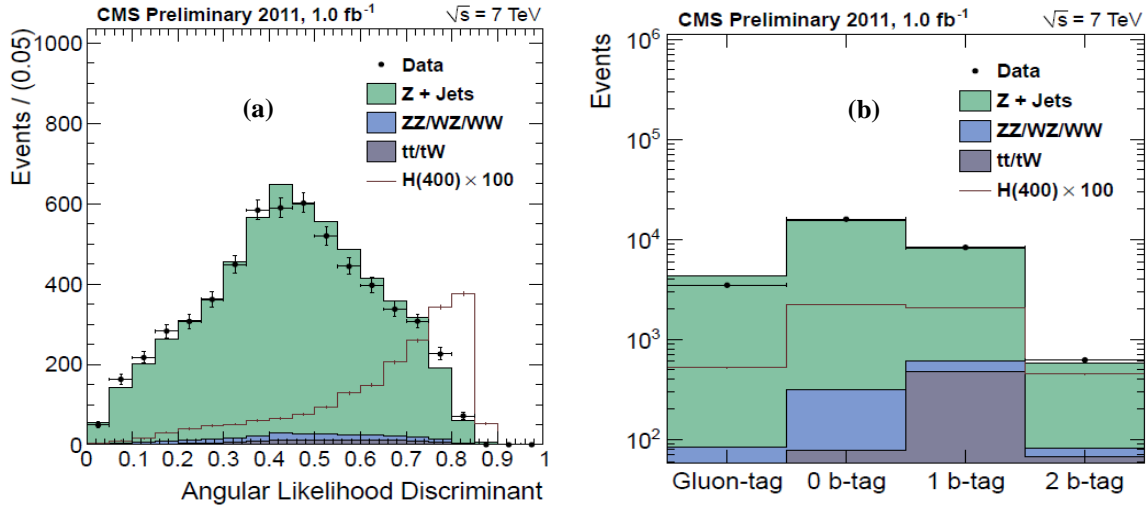

Figure 1: Distribution of the angular likelihood discriminant (a) and flavor tagging category (b), including the gluon-tagged category. Points with error bars show distributions of data after pre-selection requirements, solid histograms depict the background expectation from simulated events with the different components illustrated. Open histograms indicate the expected distribution for a Higgs boson with mass 400 GeV, multiplied by a factor of 100 for illustration.

### 3. Background Estimation

The primary discrimination power between signal and background is provided by the invariant mass distribution of the Higgs boson, $m_{2\ell 2j}$, which is used for the statistical analysis. Data containing a Higgs boson signal would have a distinct resonance peak on top of the continuum background distribution. Therefore, it is critical to have a proper estimation of the background which is estimated from the $m_{jj}$ sidebands, defined as 60 GeV < $m_{jj}$ < 75 GeV or 105 GeV < $m_{jj}$ < 130 GeV. The composition and distribution of the dominant backgrounds in the sidebands is in agreement with that in the signal region, 75 GeV < $m_{jj}$ <105 GeV. The advantage of this approach is that most of the systematic uncertainties on backgrounds cancel in the ratios, such as theoretical cross-section prediction and b-tagging efficiency. It also provides an in-situ normalization of the background and adjusts to the shape of the $m_{2\ell 2j}$ mass spectrum should there be any discrepancy between simulation and data. The above procedure is applied independently in each b-tag category since background composition varies between categories. In all cases, the dominant backgrounds include Z+jets with either light or heavy flavor jets and top background, both of which populate the $m_{jj}$ signal region and the $m_{jj}$ sidebands. The diboson background amounts to less than 5% in the zero and one b-tag categories and about 10% in the 2 b-tag category. Additional cross-checks of the dominant background contributions from Z+jets and top production have been performed using γ+jets and eμjj control samples, respectively. Figure 2 shows the observed data, the expected background, and an example of expected signal for the events with 0, 1 and 2 b-tags. Good agreement is observed between data and predictions of background distributions after the pre-selection requirements, where the additional contribution of a Higgs boson signal would be indistinguishable above the large background.

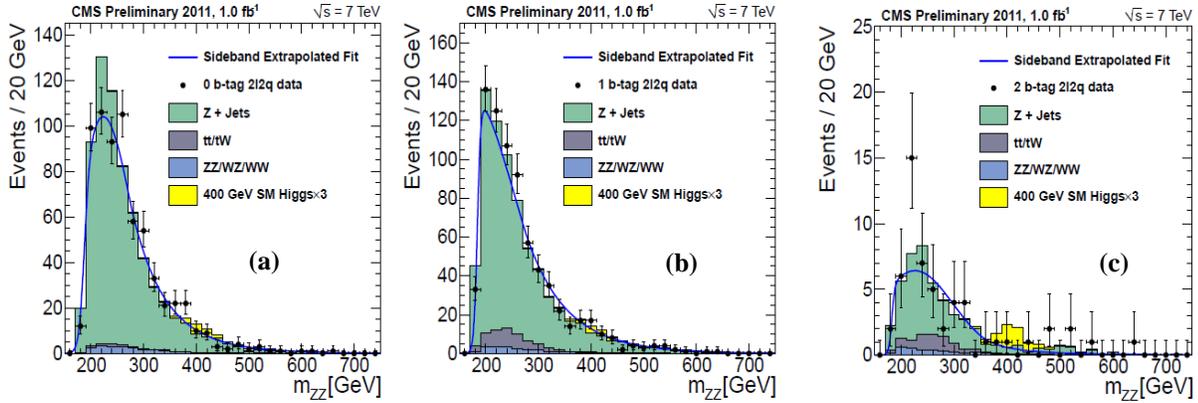

Figure 2: The $m_{2\ell 2j}$ invariant mass distribution after final selection in three categories: 0 b-tag (a), 1 b-tag (b), and 2 b-tag (c). Points with error bars show distributions in data, solid histograms depict the background expectation from simulated events with the different components illustrated. Also shown is a hypothetical signal with the mass of 400 GeV and cross section 3 times that of the SM Higgs. Solid curved line shows prediction of background from sideband extrapolation procedure.

## 4. Results

The Higgs boson search is performed by looking for an excess of data over the SM background expectation in the $m_{2\ell 2j}$ distribution. The number and distribution of candidate H→ZZ→2ℓ 2jets events observed in the data agree with the expected background within the uncertainties, with no indication of an excess. Upper limits are set on the Higgs boson production cross section relative to its predicted SM value as a function of $m_H$. The limits are extracted from a maximum likelihood fit to the $m_{2\ell 2j}$ distribution following the $CL_s$ method [11]. All systematic uncertainties are taken into account. The main sources of systematic uncertainty affecting the signal yield are the uncertainties on the total cross section and branching ratio, ~17%, and the integrated luminosity, ~6%. The uncertainty on b-tagging efficiency varies from 1% to 20% depending on the category. Other sources of systematic uncertainties from lepton energy scale, lepton reconstruction efficiency, jet resolution and efficiency, pile-up, quark-gluon discrimination, missing transverse energy and production mechanism have also been considered.

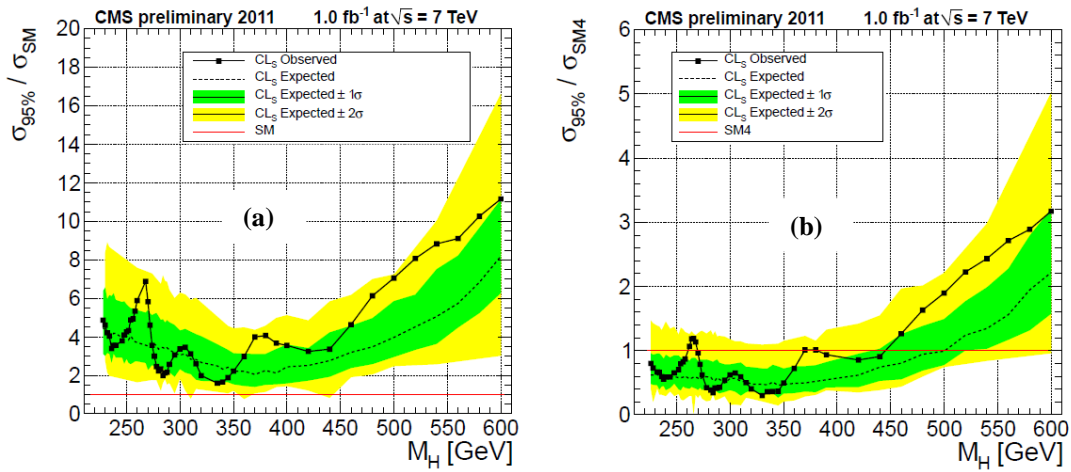

Figure 3: (a) Observed (solid) and expected (dashed) 95% CL upper limit on the ratio of the Higgs boson production cross section to the SM expectation using 1 fb$^{-1}$ of data obtained with the CLs technique. The 68% and 95% ranges of expectation are also shown with green and yellow bands. The solid line at 1 indicates SM expectation. (b) Observed (solid) and expected (dashed) 95% CL upper limit on the ratio of the Higgs boson production cross section to the expectation with the SM4 model.

Figure 3(a) shows the expected and observed limits at the 95% CL. As one moves toward the lower masses, the Z+jets background quickly overwhelms a potential signal. Cuts optimized for a higher Higgs boson mass allow one to suppress this formidable background, but eventually the signal cross section becomes too small and limits worsen. The exclusion limits in Fig. 3(a) are approaching those of the SM expectation for the Higgs boson production. A similar limit on the ratio to the Higgs boson production cross section in the SM4 model with a 4$^{th}$ generation of high mass fermions [12] is shown in Fig. 3(b). A range of SM4 Higgs mass hypotheses are excluded between 226 and 445 GeV at 95% CL, except for two windows between 261 and 270 GeV and between 370 and 381 GeV. To increase the overall experimental sensitivity to the presence of the signal, the search results obtained in such analyses focusing on different Higgs decay modes have been combined [13]. The conclusion of this combination is that the SM Higgs boson is excluded at the 95% CL in two mass ranges 149-206 GeV and 300-440 GeV, as well as several narrower intervals in between. The results from the combination also excludes the SM4 Higgs boson with a mass in the range 120-600 GeV at the 95% CL.

## 5. Summary

Results of a search for the SM Higgs boson with a mass in the range $200 < m_H < 600$ GeV decaying into two Z bosons which subsequently decay to two leptons and two quark jets has been discussed. These results are based on a data sample corresponding to an integrated luminosity of 1 fb$^{-1}$, recorded with the CMS detector at the LHC. No evidence for a SM-like Higgs boson has been found and upper limits on the production cross section for the SM Higgs boson have been set over the entire mass range considered. The analysis have also excluded a large range of Higgs mass hypotheses in the model with the fourth generation SM-like couplings of the Higgs and, when results are combined with other Higgs decay channels from CMS, excluded a wide range of masses of the SM Higgs [13].